\documentclass[conference]{IEEEtran}
\IEEEoverridecommandlockouts
\usepackage{cite}
\usepackage{amsmath,amssymb,amsfonts}
\usepackage{algorithmic}
\usepackage{graphicx}
\usepackage{textcomp}
\usepackage{xcolor}
\usepackage{caption}
\usepackage{subcaption}
\usepackage{multirow}
\usepackage{hyperref}
\usepackage[flushleft]{threeparttable}
\def\BibTeX{{\rm B\kern-.05em{\sc i\kern-.025em b}\kern-.08em
    T\kern-.1667em\lower.7ex\hbox{E}\kern-.125emX}}
\usepackage{array}
\newcolumntype{C}[1]{>{\centering\arraybackslash}m{#1}}
\usepackage{float}
\begin{document}

\title{Physical Layer Design for Ambient IoT}
\author{\IEEEauthorblockN{Rohit Singh\IEEEauthorrefmark{1},
Anil Kumar Yerrapragada\IEEEauthorrefmark{2}, Radha Krishna Ganti\IEEEauthorrefmark{3}}
\IEEEauthorblockA{Department of Electrical Engineering\\
Indian Institute of Technology
Madras \\ Chennai, India  600036\\
Email: \IEEEauthorrefmark{1}rohitsingh@smail.iitm.ac.in,
        \IEEEauthorrefmark{2}anilkumar@5gtbiitm.in,       
	\IEEEauthorrefmark{3}rganti@ee.iitm.ac.in
}
}

\makeatletter

\maketitle
\begin{abstract}
There is a growing demand for ultra low power and ultra low complexity devices for applications which require maintenance-free and battery-less operation. One way to serve such applications is through backscatter devices, which communicate using energy harvested from ambient sources such as radio waves transmitted by a reader. Traditional backscatter devices, such as RFID, are limited by range, interference, low connection density, and security issues. To address these problems, the Third Generation Partnership Project (3GPP) has started working on Ambient IoT (A-IoT). For the realization of A-IoT devices, various aspects ranging from physical layer design, to the protocol stack, to the device architecture should be standardized. In this paper, we provide an overview of the standardization efforts on the physical layer design for A-IoT devices. The various physical channels and signals are discussed, followed by link level simulations to compare the performance of various configurations of reader to device and device to reader channels.


\end{abstract}


\section{Introduction}

The Internet of Things (IoT) encompasses multiple applications such as healthcare, manufacturing, security, agriculture, smart homes, etc. to perform various tasks such as identification, tracking, sensing, communication, computation, and semantics. In supply chain management, a widely deployed technology for identification and tracking is radio frequency identification (RFID), which uses backscatter communication to reflect signals back to the transmitter. Despite its low power consumption, low cost, and small form factor, RFID suffers from critical problems such as security, interference, low range, and low connection density~\cite{1589116}.

Various other advances have been made in the field of IoT using cellular communications, focusing mainly on machine-type communication use cases. Cellular IoT technologies such as Narrow Band IoT (NB-IoT) and LTE Machine Type Communication (LTE-MTC) integrate well with the existing cellular framework, consume low power and use less time-frequency resources~\cite{abou2021nb}. 

However, to scale up the number of devices, battery-less or very low power operation is required. Backscatter communication with energy harvesting~\cite{7120024} is still the most feasible way forward in the design of battery-less devices, but these devices should also overcome the difficulties faced by RFID technology. Several studies have been conducted on the complexity and power consumption of backscatter communications~\cite{liu2013ambient}. To demonstrate that backscatter communication works, hardware prototypes without a battery are described in~\cite{liu2013ambient}. These devices are shown to achieve a data rate of $1$ Kbps over a distance of 1.5 feet and 2.5 feet in outdoor and indoor environments respectively. An overall design perspective of ambient backscatter devices, including architecture design, hardware design, and network protocols, are discussed in~\cite{8368232}.
The potential shown by ambient backscatter communication makes it a strong contender for next generation low-power cellular IoT technology. 

  3GPP has completed a RAN Working Group level study on A-IoT~\cite{TR_38_848}, followed by multiple further studies at the sub-working group level~\cite{RAN1_119_notes}. 
  A new network function in the 5G core network to manage the communication and data flow of A-IoT devices with their application server is described in~\cite{10453191}. A link budget analysis of the device to reader link is shown in~\cite{10463656}. Various challenges in the device architecture for reflection amplifier, frequency shifter, energy harvesting and harmonic interference are discussed~\cite{10704336}. Random access, energy harvesting and user plane protocol stack design are discussed in~\cite{10608245}. An initial prototype and demonstration of the feasibility of the A-IoT devices is given in~\cite{10705318}.

In this paper we focus on the following aspects: 
\begin{itemize}
    \item A summary of the A-IoT applications, device types and connection topologies.
    \item Physical layer design aspects of A-IoT being considered in 3GPP.
    \item The transmitter and receiver design aspects of A-IoT being considered in 3GPP.
    \item Random Access framework for A-IoT.
    \item Performance evaluation of different configurations of Reader to Device (R2D) and Device to Reader (D2R) links.
\end{itemize}

\section{Background on Ambient IoT in 3GPP}
This section provides the background on use cases, device types and various deployment topologies for A-IoT.

\subsection{Use Cases for Ambient IoT}
A-IoT use cases are divided into 8 categories, four cases each for indoor and outdoor. The use cases are grouped based on their deployment environments and functionality~\cite{TR_38_848} and are listed below.
\begin{enumerate}
    \item rUC1: Indoor inventory
        \begin{itemize}
            \item Automated warehousing
            \item Automobile manufacturing
            \item Airport terminal / shipping port
            \item End-to-end logistics
        \end{itemize}
    \item rUC2: Indoor sensor
        \begin{itemize}
            \item Smart homes
            \item Smart agriculture
            \item Base station machine room environmental supervision
        \end{itemize}
    \item rUC3: Indoor positioning
        \begin{itemize}
            \item Finding remote lost item
            \item Location service
            \item Ranging in a home
            \item Positioning in shopping centre
        \end{itemize}
    \item rUC4: Indoor command
        \begin{itemize}
            \item Online modification of medical instruments status
            \item Device activation and deactivation
            \item Elderly health care
        \end{itemize}
    \item rUC5: Outdoor inventory
        \begin{itemize}
            \item Medical instruments inventory management and positioning
            \item Non-public network for logistics
            \item Airport terminal / Shipping port
            \item Automated supply chain distribution
        \end{itemize}
    \item rUC6: Outdoor sensor
        \begin{itemize}
            \item Smart grids
            \item Forest fire monitoring
            \item Smart manhole cover safety monitoring
            \item Smart bridge health monitoring
        \end{itemize}
    \item rUC7: Outdoor positioning
        \begin{itemize}
            \item Finding remote lost item
            \item Location service
            \item Personal belongings finding
        \end{itemize}
    \item rUC8: Outdoor command
        \begin{itemize}
            \item Online modification of medical instruments status
            \item Device activation and deactivation 
            \item Elderly health care
            \item Controller in smart agriculture
        \end{itemize}
\end{enumerate}

\subsection{Device types}
There are 3 types of A-IoT devices, which are classified based on energy storage and independent signal generation capabilities. 
The device types are as follows.
\begin{itemize}
    \item \textbf{Device 1:} These devices have no energy storage and their transmission is based on a carrier wave using backscattering.
    \item \textbf{Device 2a:} These devices have energy storage and transmission is based on a carrier wave using backscattering. The stored energy can be used for the amplification of backscattered signals.
    \item \textbf{Device 2b:} These devices have energy storage and active RF components for independent signal generation.
\end{itemize}
Of all these three, device type 1 is the simplest and 3GPP is currently focusing on its architecture. The transmitter and receiver design, described in this paper is applicable to Device types 1 and 2a.


\subsection{Topologies}
\begin{figure}[h!]
        \centering        
        \includegraphics[width=\columnwidth]{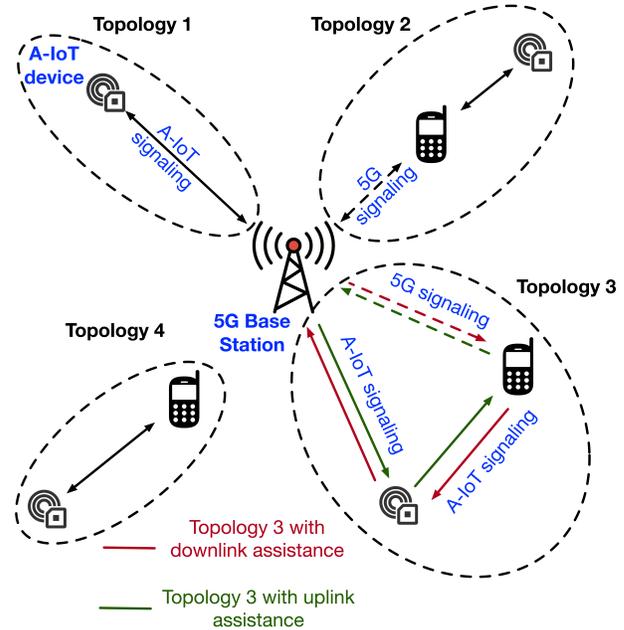}
        \caption{Various topologies for Ambient IoT}
        \label{fig:Topology}
    \end{figure}

For the evaluation of A-IoT devices, 3GPP is considering  four different topologies which are shown in Fig.~\ref{fig:Topology}.
\begin{itemize}
    \item \textbf{Topology 1:}
    In Topology 1, an A-IoT device directly communicates with a base station in a bi-directional manner. The communication between the base station and the A-IoT device includes A-IoT data and/or signaling. This topology includes the possibility that the base station transmitting to the A-IoT device is different from the base station receiving from the A-IoT device.

    \item \textbf{Topology 2:}
    In Topology 2, an A-IoT device communicates bidirectionally with an intermediate node between the device and base station. In this topology, the intermediate node can be a relay, IAB node, UE, repeater, etc. which is capable of A-IoT. The intermediate node transfers A-IoT data and/or signaling between BS and the A-IoT device.
    
    \item \textbf{Topology 3:}
    In Topology 3, an A-IoT device transmits data/signaling to a base station, and receives data/signaling from the assisting node; or the A-IoT device receives data/signaling from a base station and transmits data/signaling to the assisting node. In this topology, the assisting node can be a relay, IAB, UE, repeater, etc. which is capable of A-IoT communication.

    \item \textbf{Topology 4:}
    In Topology 4, an A-IoT device communicates bidirectionally with a UE. Communication between the UE and the A-IoT device includes A-IoT data and/or signaling. 
    
\end{itemize}

\section{Physical Layer design for Ambient IoT}
 Similar to the 5G NR standards, A-IoT also requires bi-directional communication. The Reader to Device (R2D) link is similar to  the downlink (DL) channel in 5G NR, and it contains the physical channel and signals transmitted from the Reader to the A-IoT Device. The Device to Reader (D2R) link is similar to the uplink (UL) channel in 5G NR, and it contains the physical channel and signals transmitted from the A-IoT device to the reader.



\subsection{Reader to Device Communication}
In A-IoT, the physical layer should be designed in such a way that the device architecture is simple, of low complexity and consumes low power. Because of this, only one physical channel is defined for the R2D link, which is known as the Physical Reader to Device Channel (PRDCH). This channel is used to carry the data and the control information from the reader to device. A dedicated broadcast (eg. 5G PBCH like channel) and a dedicated control channel (eg. 5G PDCCH like channel) are not supported for R2D. Physical signals (pilots) such as DMRS, CSIRS, PTRS are not supported in R2D.

\begin{figure*}[!ht]
    \centering  
    \begin{subfigure}{0.49\textwidth}
        \centering
        \includegraphics[width=0.7\textwidth]{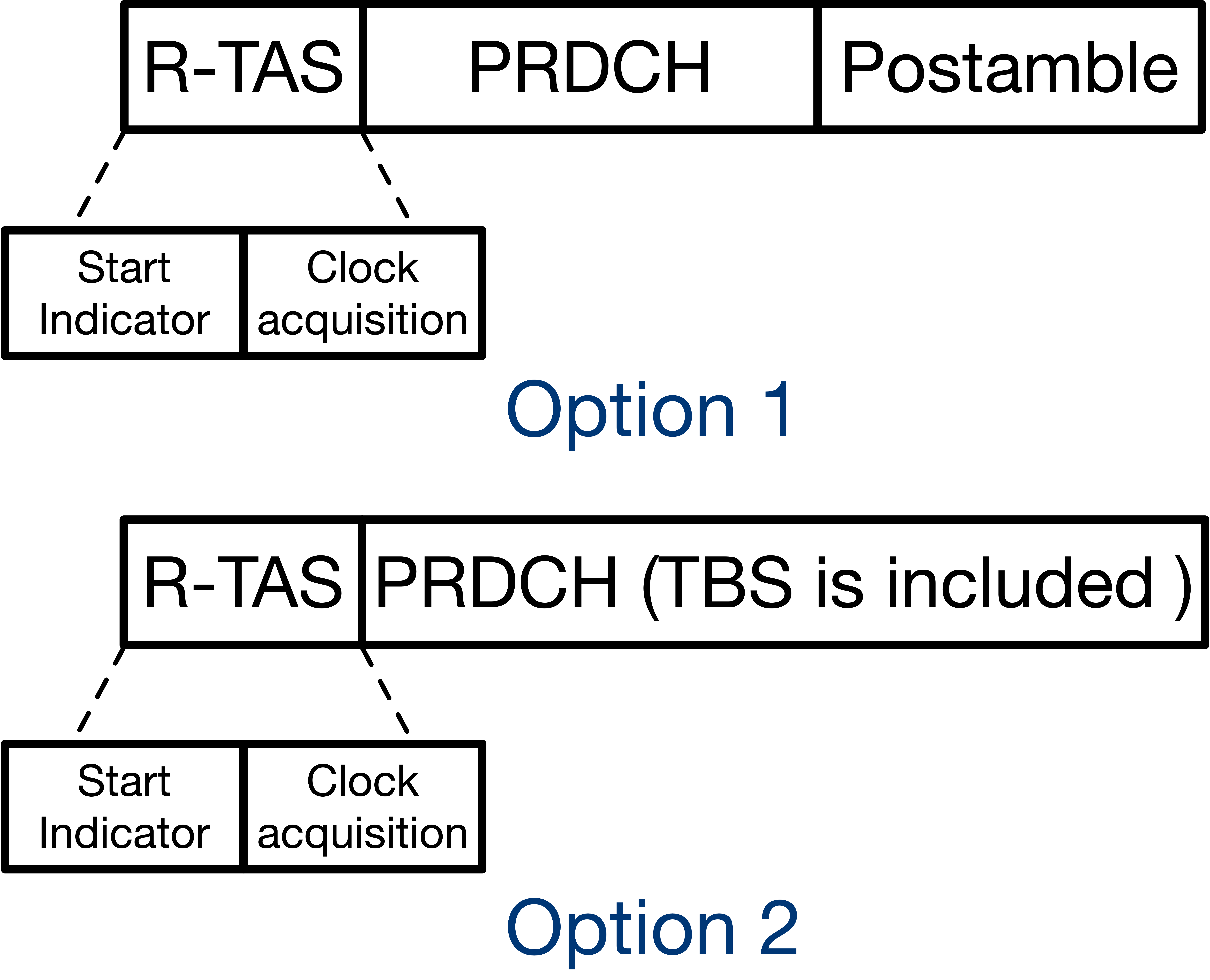}
        \caption{}
        \label{fig:R2D_timing}  
    \end{subfigure}
    \hfill
    \begin{subfigure}{0.5\textwidth}
        \centering
        \includegraphics[width=\textwidth]{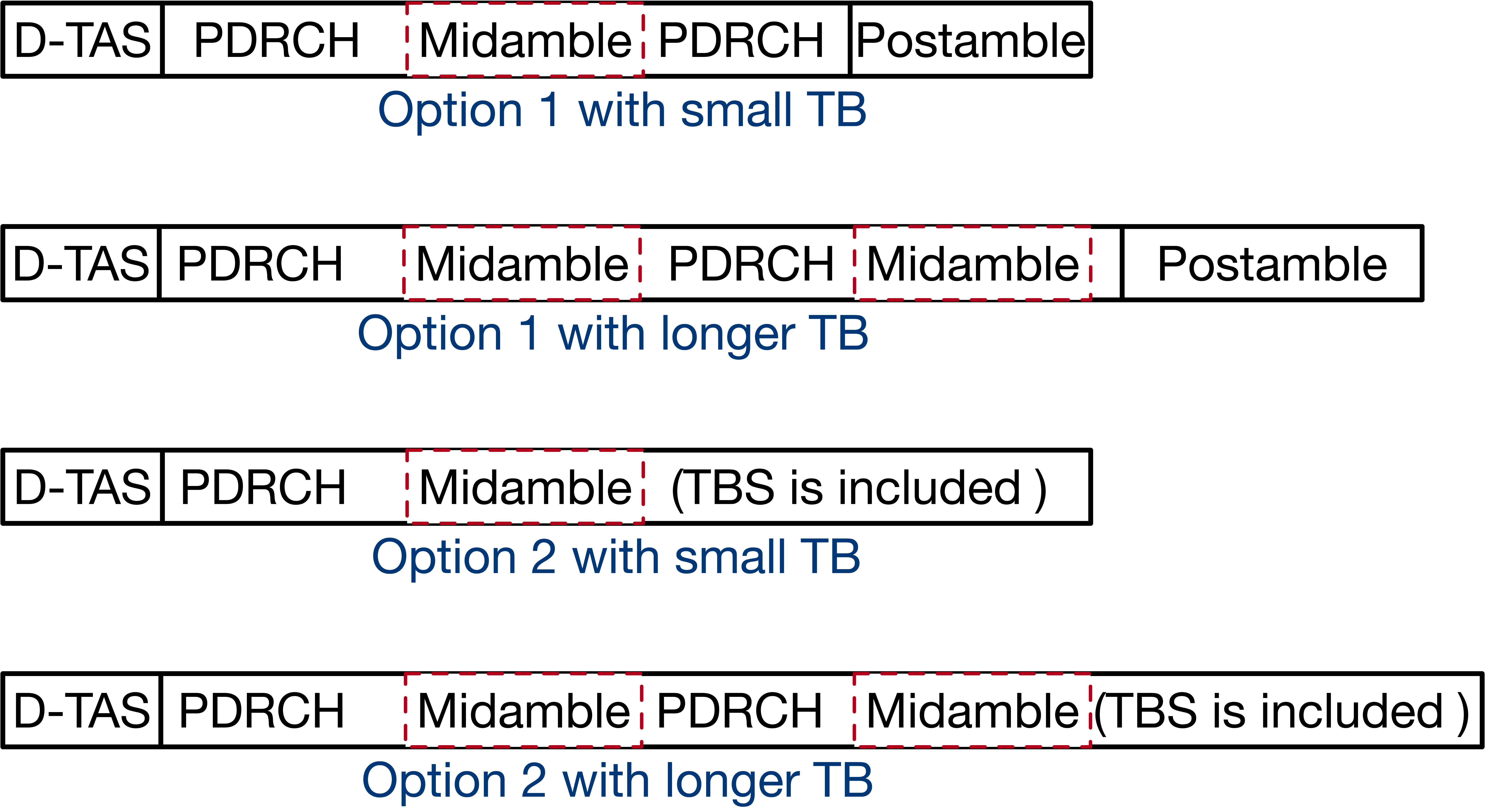}
        \caption{}
        \label{fig:D2R_timing}
    \end{subfigure}
    \caption{Physical channels and signals for A-IoT. (a) Reader to device (R2D) (b) Device to reader (D2R)}
\end{figure*}

As this is an asynchronous system, the starting and ending of the PRDCH transmission needs to be specified to the device. As shown in Fig.~\ref{fig:R2D_timing}, for initial timing and clock identification, an R2D Timing Acquisition Signal (R-TAS) is defined. This signal is transmitted just before the PRDCH and consists of two parts (which are preamble based): a start indicator signal and a clock acquisition signal. The start indicator contains an On/Off pattern and indicates the start of the PRDCH to the device. The clock acquisition signal is based on On/Off Keying (OOK) without line coding. It contains at least two falling or two rising edges, which indicates the chip duration to the device. There are two options for determining the end of a PRDCH transmission. One is to send a postamble right after the PRDCH to indicate to the device that the PRDCH transmission has ended. Another option is to send the Transport Block Size (TBS) in the control information on the PRDCH as shown in Fig.~\ref{fig:R2D_timing}.

Considering the very low power requirements of A-IoT devices and the fact that the receiver is based on RF envelope detection, OOK based waveforms are used for the PRDCH. In order for the signaling to co-exist with 5G NR, the R2D waveforms are constructed using DFT spread OFDM. This OFDM based OOK waveform can support multiple chips per OFDM symbol. OOK-1 is used for 1 chip per OFDM symbol and OOK-4 is used for multiple chips per OFDM symbol. In OOK-1, all the scheduled subcarriers in an OFDM symbol are modulated using a single chip. In OOK-4, $M$ OOK time domain chips are transformed using DFT or least squares before subcarrier mapping~\cite{TR_38_869}. From a reader (5G base station) perspective, the PRDCH transmission is similar to that of 5G NR and from the A-IoT device perspective, it is simply an OOK waveform. 

\subsubsection{PRDCH Transmitter design}
The maximum number of bits that can be transmitted in the PRDCH channel is $1000$. For PRDCH signal generation, a subcarrier spacing of $15$ KHz is used. Multiple chips can be mapped to one OFDM symbol for higher data rates. The transmitted bandwidth depends on the number of chips per OFDM symbol. The minimum number of PRBs required to transmit $M$ chips per OFDM symbol is given in Table 6.1.1.4-1 of~\cite{TR_38_769}.
    
\begin{figure*}[h]
    \centering
    \begin{subfigure}{\textwidth}
        \centering
        \includegraphics[width=\linewidth]{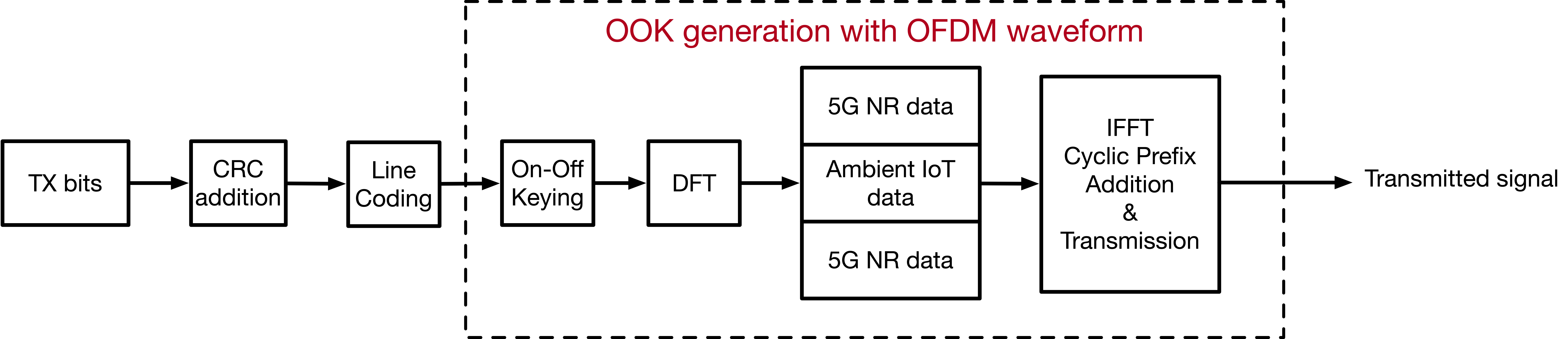} 
        \caption{}
        \label{fig:PRDCH_TX}
    \end{subfigure}
    
    \vspace{1em} 
    
    \begin{subfigure}{\textwidth}
        \centering
        \includegraphics[width=\linewidth]{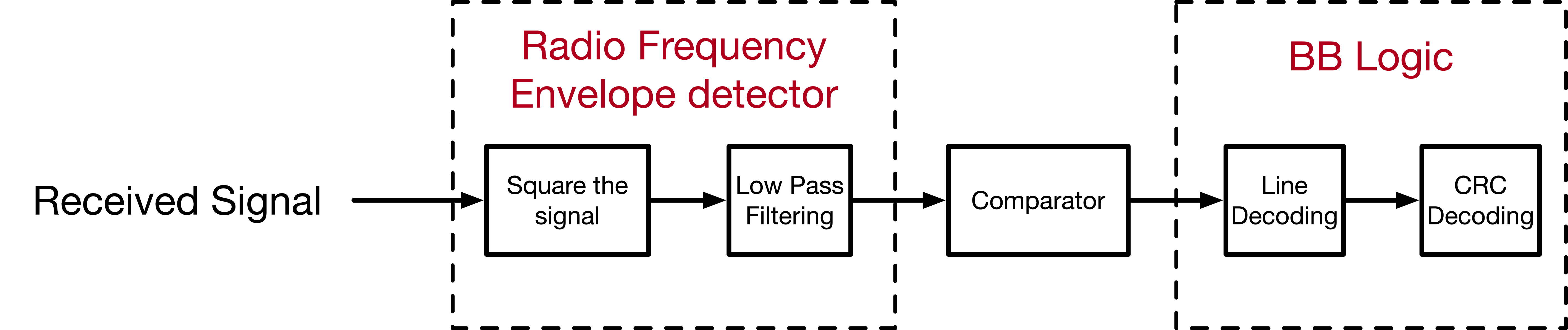}
        \caption{}
        \label{fig:PRDCH_RX}
    \end{subfigure}
    
    \caption{PRDCH transmitter and receiver design. (a) PRDCH transmitter architecture at the reader. (b) PRDCH receiver architecture at the Ambient IoT device.}
    \label{fig:PRDCH_TX_RX}
\end{figure*}

Fig.~\ref{fig:PRDCH_TX} shows the overall transmitter design of the PRDCH. First, the transport block undergoes CRC addition. The CRC length and the associated polynomial depends on the length of the transport block. Two different CRC polynomials are considered - 6 bit and 16 bit, from the existing 5G NR specification TS 38.212. 
The CRC polynomials are given by $g_{CRC16}(D) = D^{16} + D^{12} + D^{5} + 1 $ and $g_{CRC6}(D) = D^{6} + D^{5} + 1 $. After CRC attachment, line coding is performed. Two different line codes can be used -  Manchester and Pulse Interval Encoding (PIE). Each line coded symbol is called a chip. In case of Manchester encoding, bit 0 maps to chips \{10\} and bit 1 maps to chips \{01\}. In PIE, bit 1 maps to chips \{1,0\} where the duration of the high voltage chip (1) is longer than that of the low voltage chip (0). Bit 0 maps to chips \{10\} with equal duration of the high and low voltage chips. Each bit corresponds to two chips and there is a transition between the chips for each bit. Implementation-wise, Manchester is better than the PIE, since the PIE has different widths for chip 0 and chip 1, which can cause an error propagation. The transition of the chips in the line coding helps the device to achieve chip level time tracking. The chip length in the time domain depends on the SCS and the number of chips to be transmitted per OFDM symbol (M). 
\[
\text{Chip Length} = \frac{1}{M\times SCS}.
\]
Line coding is followed by OOK generation with a DFT spread OFDM waveform wherein a DFT operation is done based on the number of subcarriers available. If the number of chips is less than the available subcarriers, the chips are repeated before the DFT. Next, the output is mapped to the 5G NR resource grid, followed by guard band addition, IFFT and Cyclic Prefix (CP) addition in the time domain. Due to the different CP lengths in 5G for different OFDM symbols, the chip durations will be different and this can be handled at the receiver in one of two ways:
\begin{itemize}
    \item The device should remove the CP without the transmitter specifying the CP
    \item The reader should ensure that no falling/rising edge happens because of the CP.
\end{itemize}
The time domain data is up-converted and sent over the air. The output in the time domain follows the same pattern as the chips.

The effective datarates depend on the number of chips that are transmitted in each OFDM symbol, the number of bits transmitted, the CRC length, and the number of OFDM symbols transmitted.
\[
\text{Data Rate} = \frac{\text{Effective number of bits transmitted}} {\text{Total number of OFDM symbols}}
\]

\subsubsection{PRDCH Receiver Design}
As discussed previously, R2D has 3 parts (for option 1), which are R-TAS, PRDCH, and the postamble. R-TAS is preamble-based and is known at both the transmitter and the receiver. The start indicator part of the R-TAS provides the threshold information to the device. The end of the start indicator tells the device when the clock acquisition signal starts. The clock acquisition informs the device about the chip duration. Since A-IoT devices have a very high sampling frequency offset (SFO), up to $\sim10^5$ ppm, the clock acquisition part helps the device to correct for the SFO initially. The threshold information, the chip length, and the correct sampling frequency obtained from R-TAS helps the device to decode the PRDCH. Device 1 \& 2a support RF-ED (radio frequency - envelope detection) and device 2b supports RF-ED, IF-ED (intermediate frequency - envelope detection) and ZIF (zero intermediate frequency) receiver. 
A common implementation of the RF-ED receiver is based on squaring and passing through a low pass filter (LPF). In device types 2a and 2b, the RF-ED output is passed through a baseband amplifier and a low pass filter to increase the signal quality and filter out the high frequency components. The LPF is followed by a comparator to compare the signal with the previously deterimined threshold and convert the analog signals to bits. This is followed by the baseband digital logic where line and CRC decoding take place.


\subsection{Device to Reader Communication}
Similar to R2D, the D2R link has only one physical channel, the Physical Device to Reader Channel (PDRCH), that carries the data and control information. Due to the simple design of the device, dedicated control channel (eg. 5G PUCCH like channel), and random access channel (eg. 5G PRACH like channel) are not considered for A-IoT. Reference signals such as DMRS, PTRS, SRS are not considered. 

Due to the asynchronous nature of A-IoT communication, timing signals should be transmitted from the device to the reader. As shown in Fig.~\ref{fig:D2R_timing}, a preamble-based D2R timing acquisition signal (D-TAS) is transmitted before the PDRCH to indicate the starting point of the PDRCH. This signal can also be used for SFO, CFO and channel estimation. The D-TAS signal is a binary signal that has good auto-correlation and cross-correlation properties. For D2R channel estimation, CFO estimation, time tracking, etc., a midamble signal is used. The number of midambles used depends on the size of the PDRCH data, but there is a trade off between the number of midambles and achievable datarates. For smaller TB sizes, one midamble is sufficient, and for larger TB sizes multiple midambles can be used for better estimation of SFO, CFO and channel state. Similar to the R2D link, there are two options to indicate the end of a PDRCH transmission. One is to send a postamble at the end of the PDRCH transmission and another is to indicate the Transport Block Size (TBS) as control information on the PDRCH. 




\subsubsection{PDRCH Transmitter Design}
PDRCH carries the D2R data and control information. A non OFDM signal is used and the device generates data in the baseband and modulates it onto a carrier wave (CW). Device 1 and 2a use a provided carrier wave to backscatter the data. Device 2b has an independent carrier wave generation. In A-IoT, there are multiple ways of providing the carrier wave to the device.

\begin{figure*}[h]
    \centering
    \begin{subfigure}{\textwidth}
        \centering
        \includegraphics[width=\linewidth]{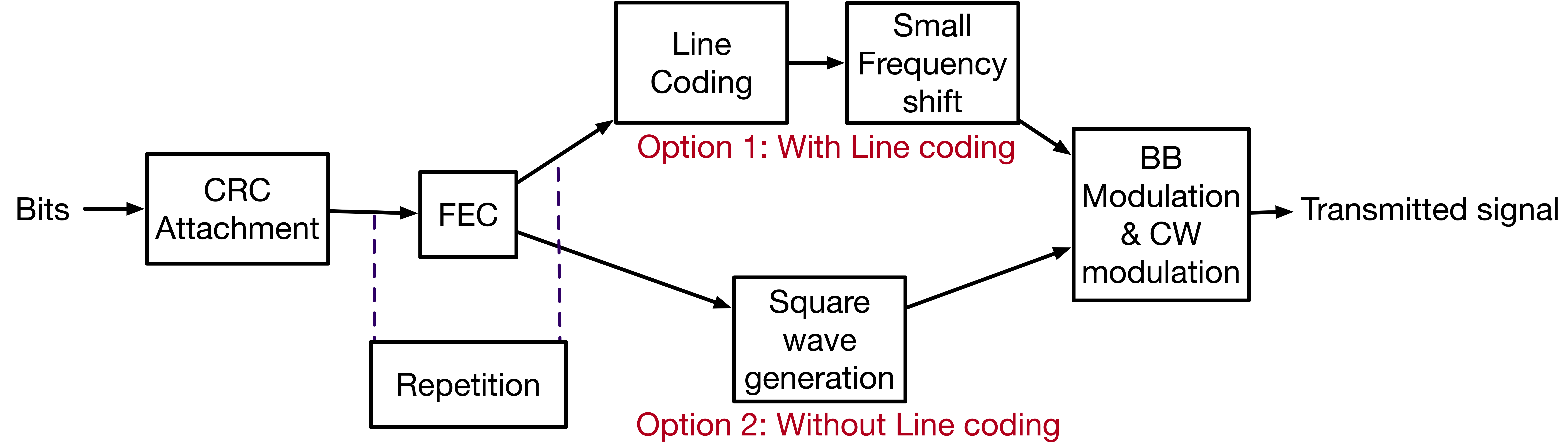} 
        \caption{}
        \label{fig:PDRCH_TX}
    \end{subfigure}
    
    \vspace{1em} 
    
    \begin{subfigure}{\textwidth}
        \centering
        \includegraphics[width=\linewidth]{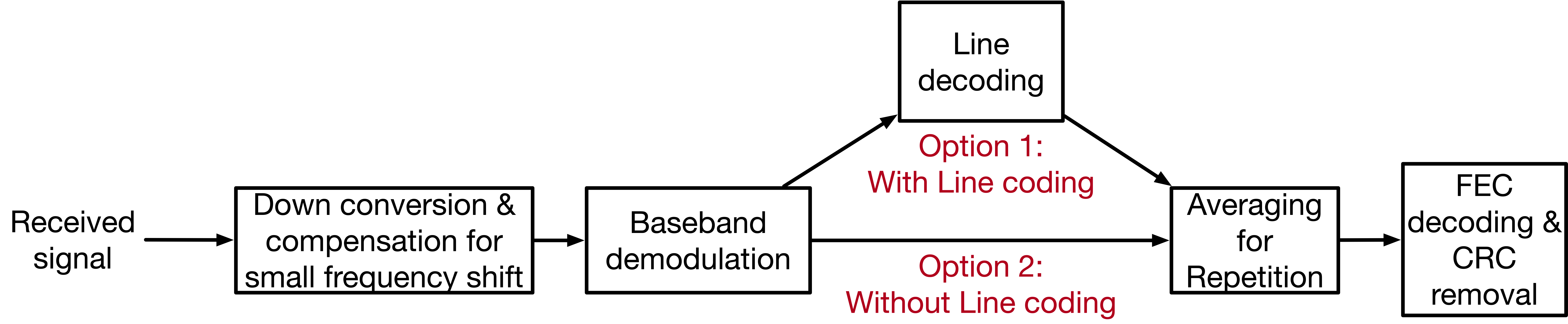}
        \caption{}
        \label{fig:PDRCH_RX}
    \end{subfigure}
    
    \caption{PDRCH transmitter and receiver design. (a) PDRCH transmitter architecture at the Ambient IoT device. (b) PDRCH receiver architecture at the reader.}
    \label{fig:PDRCH_TX_RX}
\end{figure*}

For topology 1, the following three cases are supported for the CW transmission~\cite{TR_38_769}:
\begin{itemize}
    \item \textbf{Case 1-1:} CW is transmitted from inside the topology, in the DL spectrum
    \item \textbf{Case 1-2:} CW is transmitted from inside the topology, in the UL spectrum
    \item \textbf{Case 1-4:} CW is transmitted from outside the topology, in the UL spectrum
\end{itemize}
For topology 2, the supported carrier wave cases are as follows:
\begin{itemize}
    \item \textbf{Case 2-2:} CW is transmitted from inside the topology (i.e., intermediate UE), in the UL spectrum
    \item \textbf{Case 2-3:} CW is transmitted from outside the topology in the DL spectrum 
    \item \textbf{Case 2-4:} CW is transmitted from outside the topology in the UL spectrum
\end{itemize}
 Bit-level, block-level, chip-level repetitions, and forward error correction (FEC) are used to make the communication link more robust at low power, enhancing the D2R coverage. The scheduling information required for the transmission is provided by the PRDCH and contains at least the time domain resources, frequency domain resources, modulation, coding rate, chip duration, device ID, repetition and midamble information. Fig.~\ref{fig:PDRCH_TX} shows the architecture of the PDRCH transmitter.
 
The PDRCH supports a maximum TBS of 1000 bits. Initially, the CRC is appended to the transport block, using the same polynomial and length as that of the PRDCH. Based on repetition information received in the PRDCH, the bits can be repeated after CRC addition. After the CRC addition (and repetition), the bits undergo FEC where LTE convolutional codes with constraint lengths of $4, 6, 7$ and  $8$, and coding rates of $\frac{1}{6}$, $\frac{1}{4}$, $\frac{1}{3}$ and $\frac{1}{2}$ are used. The encoded bits can undergo repetition based on the scheduling information provided by the reader. Line coding is optional in the PDRCH. If it is used, one of Manchester, FM0 and Miller codes is used. The generation of Manchester codes is similar to that of the R2D. The generation of FM0 and Miller codes are given in~\cite{epc_gen2}. To achieve multiple access of various devices, a small frequency shift is used in the baseband after line coding. If Manchester coding is used, there are two methods of achieving the small frequency shift. 

Option 1: Each Manchester codeword is repeated $R$ times, within the bit duration $T_b$, where $R = \frac{T
_b}{2\cdot \text{chip length}}$. This generates a frequency shift of $\frac{R}{T_b} = \frac{1}{2\cdot \text{chip length}} \text{Hz}$.

Option 2: Each Manchester codeword is multiplied by a square wave, where the period of each square wave is $\frac{T_b}{R}$, and $R = \frac{T_b}{2\cdot \text{chip length}}$. This generates a frequency shift of $\frac{R}{T_b} = \frac{1}{2\cdot \text{chip length}} \text{Hz}$.  
The multiplication operation is performed either as an XOR or XNOR operation between a Manchester codeword corresponding to the information bit and the square wave. If line coding is not used, the output of the FEC (after repetition) uses a square wave to achieve the small frequency shift. To achieve a shift of $\frac{R}{T_b} \text{Hz}$, each bit duration $T_b$ should include $R$ square wave periods. Each square consists of one high and one low voltage.



Small frequency shift is followed by the baseband modulation for which OOK, BPSK and MSK are supported. The baseband modulated signal is transmitted by modulating the baseband signal onto the carrier wave. The amplitude or the phase of the carrier wave are modified according to the baseband signal and sent over the air.
 
\subsubsection{PDRCH Receiver Design}
Fig.~\ref{fig:PDRCH_RX} shows the architecture of the PDRCH receiver. At the reader side, the receiver first estimates the start of the PDRCH using the correlation of the received D-TAS signal, which is preamble based. Since D2R transmission uses a carrier wave, the receiver down converts the received signal and compensates for the small frequency shift based on the shift information previously indicated to the device by the reader.
To enhance receiver performance, channel estimation and equalization can also be performed using the midamble. The baseband signal is sampled and converted to chips depending on the chip duration (for option 1 in Fig.~\ref{fig:PDRCH_TX}) or bits directly (for option 2 in Fig.~\ref{fig:PDRCH_TX}). The duration of the chip depends on the transmit frequency used by the device. In a non-coherent receiver, a threshold is used to estimate the chip or bits from the baseband data. To improve performance, adaptive thresholding can be used to remove the channel effects. Line decoding is performed, if applicable. This is followed by handling the repetition used by the device, which is done by averaging, majority voting, or soft decision decoding. To recover the transport block, the output is sent, after repetition, for FEC decoding and CRC removal.

\subsection{Random Access in Ambient IoT}
Random Access allows the A-IoT device to identify itself to the reader. In A-IoT, the random access procedure is triggered by the reader with a paging message. A reader can page a single device, a group of A-IoT devices or all the devices under its coverage. A device can receive the paging as long as it has harvested enough energy. The reader can indicate to the device whether to use contention-based or contention-free random access.  The device responds to the paging message with a D2R transmission containing a 16-bit random ID. This step is the A-IoT Msg 1. If contention-free random access is indicated by the reader, the device transmits the A-IoT Msg 1 in the allocated resources. If contention-based random access is chosen, the device chooses a random 'access occasion' for the A-IoT Msg 1. The probability of multiple devices choosing the same random ID and access occasion is low and in such cases, the device can send both the random ID and D2R data also in A-IoT Msg 1, thus reducing the number of messages between the device and the reader. The reader responds to the A-IoT Msg 1 with the A-IoT Msg 2 reflecting the random ID sent by the device. If the random ID matches, then the device assumes that the contention resolution is successful. If the random access fails, it is again triggered by the reader, not the device. The reader can also schedule further D2R data transmission using the A-IoT Msg 2.

\section{Link level simulations}

\begin{table*}[ht!]
    \centering
    \caption{Link level simulation parameters for PRDCH and PDRCH}
    \begin{tabular}{|p{3.98cm}|p{3.98cm}|p{3.98cm}|p{3.98cm}|} \hline 
         \textbf{R2D Parameter}&  \textbf{Value}  & \textbf{D2R Parameter }& \textbf{Value}\\ \hline
         Carrier frequency &0.9 Ghz&  Carrier frequency&0.9 GHz\\ \hline 
         Number of TX chains (Reader side)& 1 & Number of TX chains and antennas (Device side)&1\\ \hline 
         Subcarrier spacing&15 kHz& Subcarrier spacing&15 kHz\\ \hline 
         Number of TX antennas (Reader side)& 2 & Number of receiver antennas and chains&1, 2\\ \hline 
         Channel model&TDL-A, Delay spread = $30$ ns, Device velocity = $3$ kmph& Channel model&TDL-A, Delay spread = $30$ ns, Device velocity = $3$ kmph\\ \hline 
         Message size& 20,96 & Message size&20,96 \\ \hline 
         Sampling frequency at the device& 1.92 MSPS & Sampling frequency at the reader&1.92 MSPS\\ \hline
         Waveform & OOK-4 using DFT-s-OFDM & &\\ \hline 
         Device Type &1 & Device Type &1\\\hline
         TX bandwidth& 180 KHz (12 subcarriers) & Transmission bandwidth, Chip rate&15 kHz, 7.5 kchips/sec\\ \hline 
         Receiver type &RF-ED with fixed and adaptive thresholding & Receiver type &Non coherent with adaptive thresholding\\\hline
         Number of chips per OFDM symbol &1,2 \& 4 & Line coding, Modulation&Manchester, OOK\\\hline
    \end{tabular}
    
    \label{tab:LLS_parameters}
\end{table*}

To understand the performance of the R2D and D2R links under various Signal-to-Noise Ratios (SNR), we performed link level simulations (LLS) in MATLAB. We assume that the device has completed the Random Access procedure as described above. The various parameters that are used for transmission and reception are in accordance with the A-IoT evaluation assumptions agreements in 3GPP Release 19.

For R2D transmission, 20 and 96 TB sizes are chosen. The CRC is appended to the TB using the polynomials described above. Manchester line coding is considered. OOK-4 waveform using DFT-spread-OFDM is used since it can handle multiple chips per OFDM symbol. The number of chips per OFDM symbol are taken to be 1, 2 and 4. One PRB is allocated for the A-IoT R2D transmission. An IFFT size of 4096 is considered, which is common in a 5G base station. A SCS of $15$ KHz is considered, resulting in a sampling rate of $61.44$ Msps. The transmit data is passed through a TDL-A channel with a delay spread of $30$ ns, and a device velocity of $3$ Km/h is assumed. Two transmit antennas are used. The various parameters used for the LLS simulations of the R2D link are summarized in Table~\ref{tab:LLS_parameters}. 

\begin{figure}[ht!]
    \centering
        \begin{subfigure}{0.48\textwidth}
        \centering
        \includegraphics[width=\textwidth]{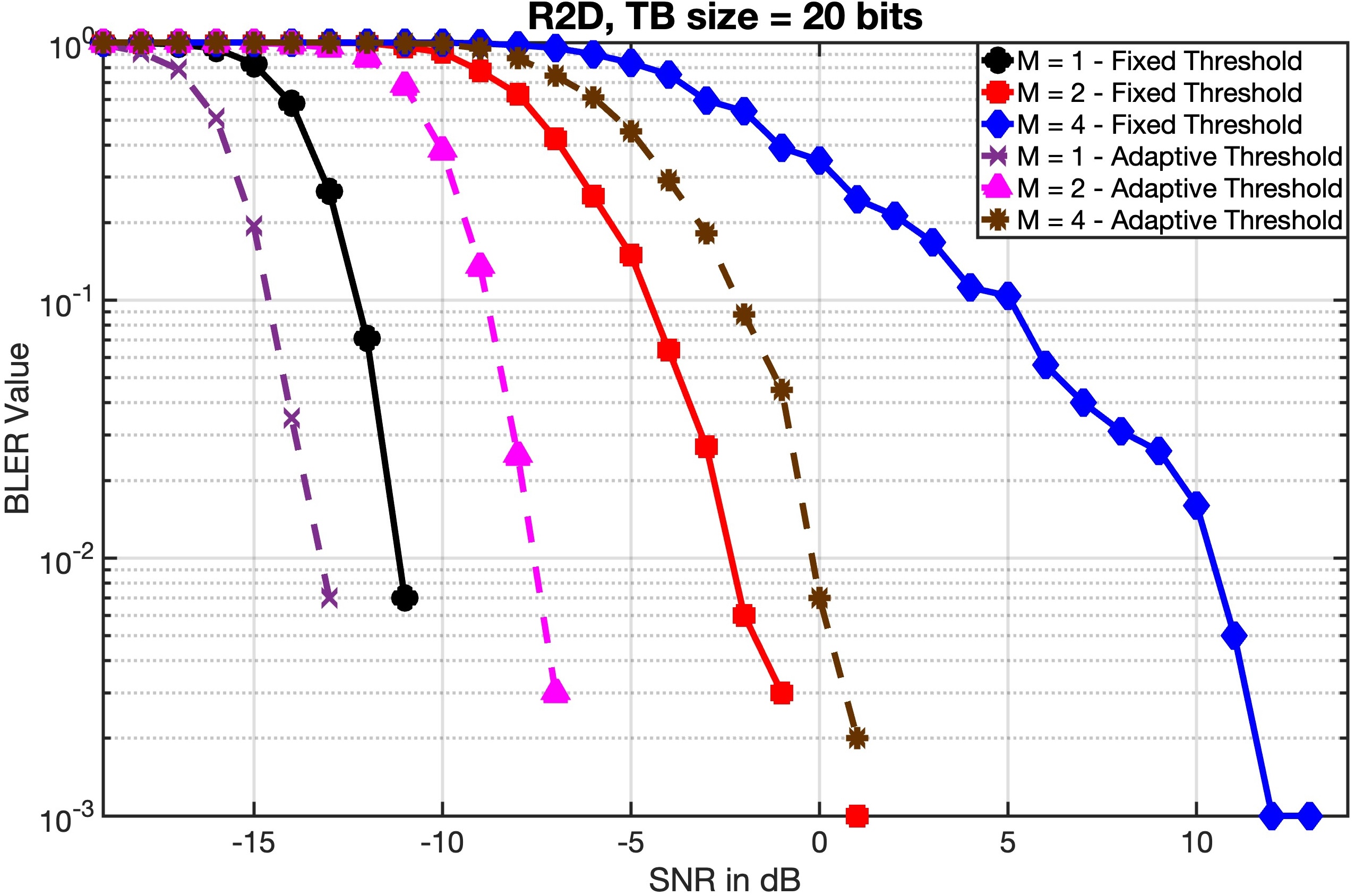}
        \caption{}
        \label{fig:R2D_lls_result_20_bits}
    \end{subfigure}
    
    \begin{subfigure}{0.48\textwidth}
        \centering
        \includegraphics[width=\textwidth]{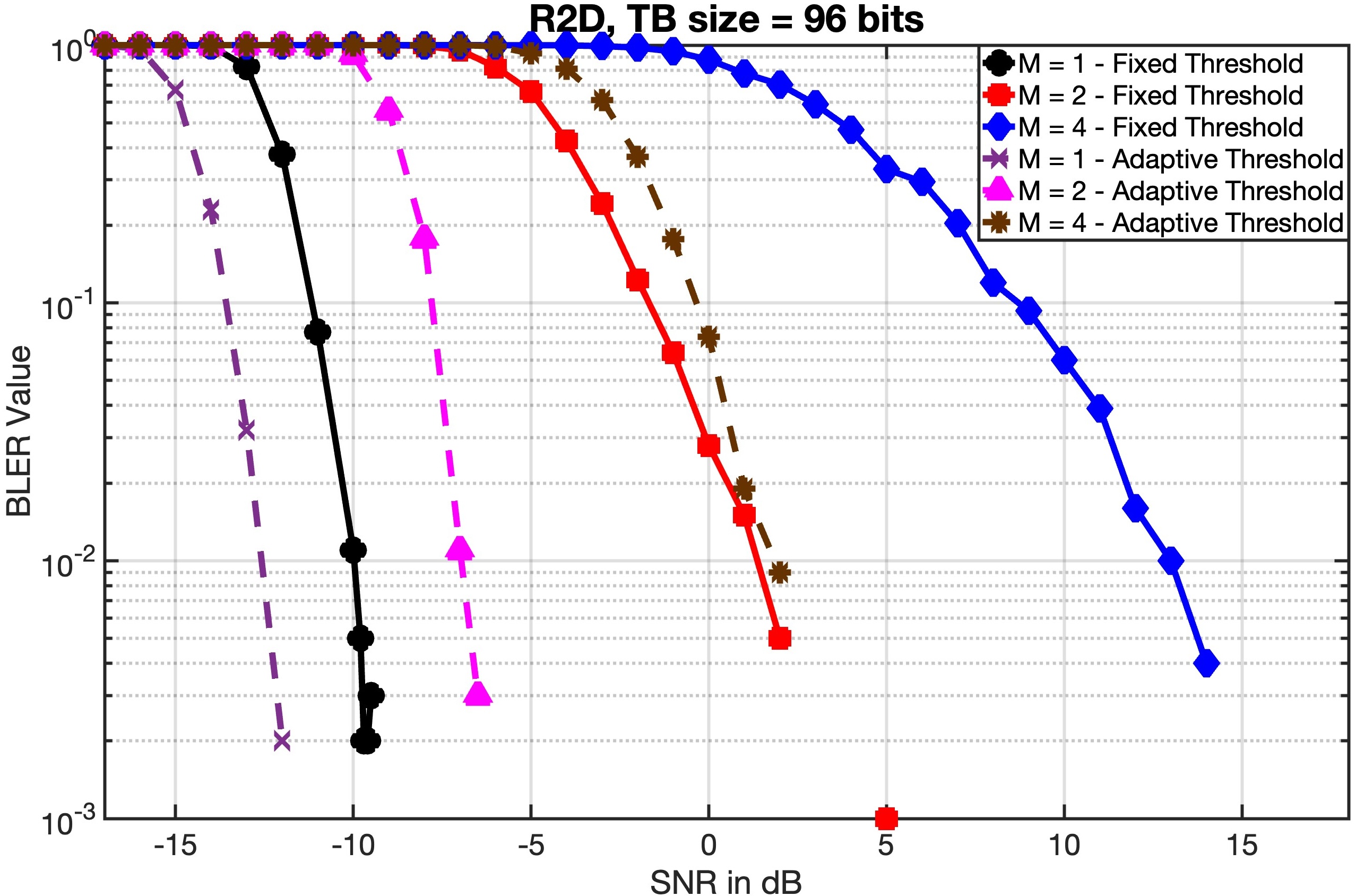}
        \caption{}
        \label{fig:R2D_lls_result_96_bits}
    \end{subfigure}
    
    \begin{subfigure}{0.48\textwidth}
        \centering
        \includegraphics[width=\linewidth]{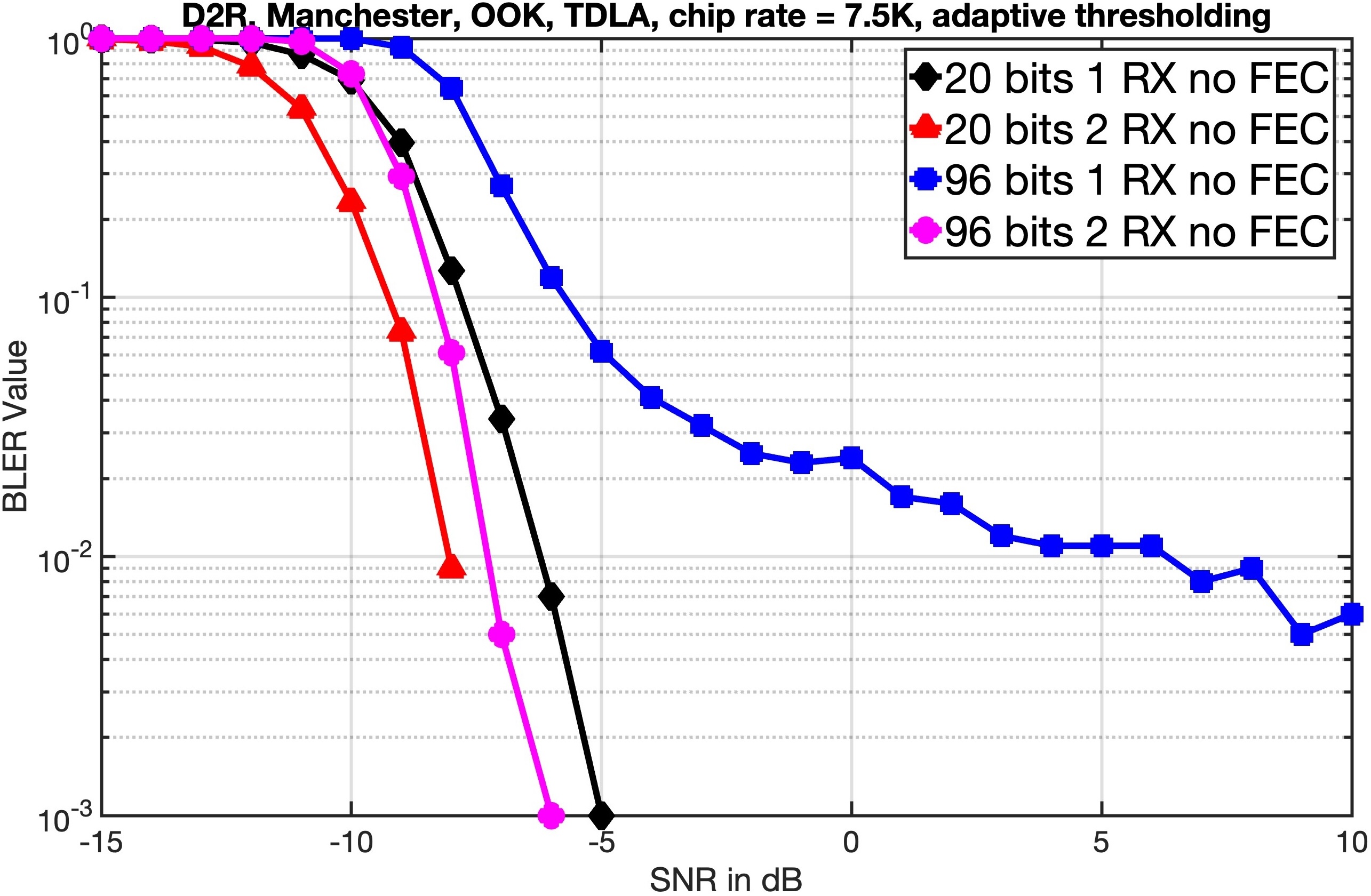}
        \caption{}
        \label{fig:D2R_lls_result}
    \end{subfigure}
    
    \caption{Link level simulations for R2D and D2R. (a) PRDCH BLER curves for TB size of 20 bits for different chip lengths and thresholding methods. (b) PRDCH BLER curves for TB size of 96 bits for different chip lengths and thresholding methods. (c) PDRCH BLER curves for various receiver antennas.}
    \label{fig:LLS_results}
\end{figure}

At the receiver, RF-ED is used. This type of receiver is suitable for all device types. The received data is resampled at a rate of $1.92$ Msps, resulting in $\frac{4096}{32M}$ samples per chip which can be combined for better receiver performance. In these experiments, two types of combining is performed. One is an average over multiple samples for each chip and comparing it with the threshold. Another is comparing each sample against the threshold and combining at the bit level using majority voting. Taking an average requires more resources and cannot be used for devices 1 and 2a. For low powered devices (1 and 2a), each sample is compared against the threshold and a majority decision is made. This can be done in the baseband logic unit of the receiver. The threshold and chip length are estimated at the receiver using the R-TAS signal. For device type 2b, taking a mean over the samples for one chip improves the receiver performance. An adaptive thresholding method with a mean taken over four consecutive chips is used to compare the chip levels. This method performs better than fixed thresholding since the threshold changes according to the channel envelope caused by Doppler. 
Fig.~\ref{fig:R2D_lls_result_20_bits} compares the Block Error Rate (BLER) performance of the R2D link with fixed thresholding and adaptive thresholding for a TB size of 20 bits. Fig.~\ref{fig:R2D_lls_result_96_bits} compares the same for 96 bits. 
We observe that adaptive thresholding performs better than fixed thresholding for any TBS. This is because adaptive thresholding acts as a kind of channel equalizer. As expected, smaller TBS shows better performance because there is a smaller variation in the channel for the smaller TBS. However, we note that the performance gain due to adaptive thresholding is higher for TB size of 96 than for 20 bits. This is because large-scale fading effects, which are prominent for larger TB sizes, are equalized with adaptive thresholding. Lastly, we observe that as the M value increases, the data rates increase, but the number of subcarriers per chip decreases, decreasing the power for that chip. Consequently, more SNR is required to achieve the same BLER for increased M.

Fig.~\ref{fig:D2R_lls_result} shows the LLS results for the various D2R cases. TB sizes of 20 and 96 bits are chosen. CRC is appended as discussed above. For a basic understanding of the link, FEC and repetition are not used for the simulations but Manchester line coding is used. For baseband modulation, OOK is used. For simplicity, these simulations are carried out in the baseband and carrier wave and small frequency shift are not used. A chip rate of $7.5$ Kchips/sec is used. TDL channel parameters are the same as the R2D LLS simulations. The data is transmitted using one transmit antenna and received using one and two antennas. In the case of two receiver antennas, equal gain combining (EGC) is performed before baseband demodulation. A sampling rate of $1.92$Msps is used at the receiver, resulting in $\frac{1.92\times10^6}{7.5\times10^3} = 256$ samples per chip at the receiver. These samples can be used for noise averaging. A non coherent receiver is used and a mean taken over every 4 chips is used as a thresholding method to detect the bits. As a baseline, for R2D, the midamble is not considered in our simulations. We observe almost $2$ dB improvement with 2 receiver antennas. Similar to R2D, the performance degrades for longer transport blocks because of the presene of Doppler. Even in the high SNR case, when 96 bits are transmitted and 1 receiver antenna is used, the receiver performance saturates because of the Doppler. The effect of the Doppler can be minimized by combining two antennas. 

\section{Conclusion}
Ambient IoT is one of the key technologies which 3GPP is currently standardizing for future releases of 5G. In this paper, we discussed the A-IoT physical layer design including the R2D and D2R links. We presented the transmitter and receiver architectures on the device and the reader side. The random access procedure of the device is also discussed at a high level. We presented and compared various configurations of PRDCH and PDRCH using link level simulations. Further works might focus on further performance enhancement (through the inclusion of repetition, FEC, channel estimation), hardware prototypes, and multi-user scenarios.


\section*{Acknowledgment}
The authors would like to thank the Ministry of Electronics and Information Technology (MeitY) for funding this work through the project "Next-Generation Wireless Research and Standardization on 5G and Beyond".


\bibliographystyle{IEEEtran}

\bibliography{bibfile}
\end{document}